# Oblivious Collaboration

Yehuda Afek [*]   Yakov Babichenko [†]   Uriel Feige [‡]   Eli Gafni [§]   Nati Linial [¶]

Benny Sudakov [‖]

May 4, 2018

Communication is a crucial ingredient in every kind of collaborative work. But what is the least possible amount of communication required for a given task? We formalize this question by introducing a new framework for distributed computation, called *oblivious protocols*.

We investigate the power of this model by considering two concrete examples, the *musical chairs* task $MC(n, m)$ and the well-known *Renaming* problem. The $MC(n, m)$ game is played by $n$ players (processors) with $m$ chairs. Players can *occupy* chairs, and the game terminates as soon as each player occupies a unique chair. Thus we say that player $P$ is *in conflict* if some other player $Q$ is occupying the same chair, i.e., termination means there are no conflicts. By known results from distributed computing, if $m \leq 2n - 2$, no strategy of the players can guarantee termination. However, there is a protocol with $m = 2n - 1$ chairs that always terminates. Here we consider an oblivious protocol where in every time step the only communication is this: an adversarial *scheduler* chooses an arbitrary nonempty set of players, and for each of them provides only one bit of information, specifying whether the player is currently in conflict or not. A player notified not to be in conflict halts and never changes its chair, whereas a player notified to be in conflict changes its chair according to its deterministic program. Remarkably, even with this minimal communication termination can be guaranteed with only $m = 2n - 1$ chairs. Likewise, we obtain an oblivious protocol for the Renaming problem whose name-space is small as that of the optimal nonoblivious distributed protocol.

Other aspects suggest themselves, such as the efficiency (program length) of our protocols. We make substantial progress here as well, though many interesting questions remain open.

---

[*]The Blavatnik School of Computer Science, Tel-Aviv University, Israel 69978. afek@tau.ac.il

[†]Department of Mathematics, Hebrew University, Jerusalem 91904, Israel yak@math.huji.ac.il

[‡]Department of Computer Science and Applied Mathematics Weizmann Institute of Science Rehovot 76100, Israel. uriel.feige@weizmann.ac.il. The author holds the Lawrence G. Horowitz Professorial Chair at the Weizmann Institute. Work supported in part by The Israel Science Foundation (grant No. 873/08).

[§]Computer Science Department, Univ. of California, LA, CA 95024, eli@cs.ucla.edu

[¶]School of Computer Science and Engineering, Hebrew University, Jerusalem 91904, Israel nati@cs.huji.ac.il

[‖]Department of Mathematics, UCLA, Los Angeles, CA, 90095. bsudakov@math.ucla.edu. Research supported in part by NSF CAREER award DMS-0812005 and by a USA-Israeli BSF grant

# 1 Introduction

In every distributed algorithm each processor must occasionally observe the activities of other processors. This can be realized by explicit communication primitives (such as by reading the messages that other processors send, or by inspecting some publicly accessible memory cell into which they write), or by sensing an effect on the environment due to the actions of other processors (such as in Carrier Sense Multiple Access channels with Collision Detection, CSMA/CD). Here we consider two severe limitations on the processors' behavior and ask how this affects the system's computational power: (i) A processor can only post a proposal for its own output, (ii) Each processor is "blindfolded" and is only occasionally provided with the least possible amount of information, namely a single bit that indicates whether its current state is "good" or "bad". Here "bad/good" stands for whether or not this state conflicts with the global-state desired by the processor. Moreover, we also impose the requirement that algorithms are deterministic (use no randomization). This new minimalist model, properly defined, is called the *oblivious model*. This model might appear to be significantly weaker than other (deterministic) models studied in distributed computing. Yet, we show that two natural problems in this field, *renaming* [1, 2] and *musical chairs* [9], can be solved optimally within the highly limited oblivious model. Furthermore, we discuss the efficiency of oblivious solutions and the relations between the oblivious model and the read/write model which is a thoroughly studied model in distributed computing [14].

The *oblivious* model can be described and formalized in two different ways: (i) in terms of the operations available to individual processors, or (ii) in terms of an oblivious oracle (as in the abstract). In either case, associated with every state of a participating processor is a proposed output, so that the state at which a processor halts thus defines its final output. In our model, an oracle mediates between the processors. The only way a processor can sense its environment is by querying the oracle about a single predicate on the current vector of outputs of the processors. Based on the single bit answer the processor needs to either halt with its current output, or proceed with its computation and propose a new output. But how can a processor's computation proceed? It has no information about the state of other processors (beyond the one bit that tells it that it must proceed), and we are forbidding randomization. Consequently, a processor's proposed output can depend only on its current state, and therefore the sequence of states that processor $p_i$ traverses is simply an infinite word $\pi_i$ over the alphabet of possible outputs. Upon receiving a negative answer from the oracle, processor $p_i$ in state $\pi_i[k]$ moves to state $\pi_i[k+1]$. Given the definition of a computational task, it is up to the programmer to design the words $\pi_i$ and the query that each processor poses to the oracle under which that task is always realized properly. Our only assumption is that the oracle correctly answers the queries, and a processor eventually halts/proceeds to the next state in his word upon a bad/good response from the oracle.

The *Musical Chairs*, $MC(n, m)$ task involves $n$ *processors* $p_1, \ldots, p_n$ and, $m$ *chairs* numbered $1, \ldots, m$. Each processor $p_i$ starts in an arbitrary chair, dictated by the input. If the input chairs are all unique, all processors are good and the input is the output. If not all input chairs are unique, the task calls for each processor to capture a chair in exclusion.

The *Renaming*$(n, m)$ task is a close relative of $MC(n, m)$. There are $m$ slots (chairs) numbered $1, \ldots, m$ and each participant has to capture a slot in exclusion. The processors have no input. If only $k < n$ processors participate, then each has to capture (output) a unique slot from the first $\min(2k-1, m)$ slots. If all the $n$ processors participate then they each capture one of the $m$ slots in exclusion.

In Section 2 we define the oblivious model in detail. For the MC and the renaming problems we use the *collision* query - a processor is good iff it is the only one to propose the current chair.

We show that in this case the general oblivious model simplifies considerably. These simplifications later help us produce an optimal solution.

Remarkably, for each processor we produce a program which is a single cyclic word on an alphabet of chairs. Furthermore, for the MC task the program can be started at any chair in the word. This provides for self stabilization [5, 6]. Namely, consider a system state where each processor occupies an exclusive chair and there are no conflicts. Suppose that the system gets perturbed, and program counters change arbitrarily. This may create conflicts, but the system will nevertheless resettle obliviously in finite time into a conflict-free safe state.

Here are the main contributions of the present paper:

1. Introduction of the general oblivious model and its specialization to the problems at hand.

2. A proof that there are tasks that are solvable in a read/write wait-free manner, but not solvable obliviously.

3. Characterization of the minimal $m$ for which there is an $MC(n, m)$ oblivious algorithm:

   **Theorem 1** *There is an oblivious $MC(n, m)$ algorithm if and only if $m \geq 2n - 1$.*

   Moreover, for all $N > n$ there exist $N$ words on $m$ chairs such that any $n$ out of the $N$ words constitute an oblivious $MC(n, 2n - 1)$ algorithm.

4. Likewise, for the Renaming problem

   **Theorem 2** *There is an oblivious $Renaming(n, m)$ algorithm if and only if $m \geq 2n - 1$.*

5. A lower bound on the number of chairs required in the oblivious MC task is derived by reduction from the renaming task, which in turn is derived from the read/write wait-free model.

6. The words in Theorem 1 use the least number of chairs, namely $m = 2n - 1$. However, the lengths of these words grows doubly exponentially in $n$. Are there oblivious MC algorithms with much shorter words? Even length $O(n)$? Perhaps even length $m$? How long can the scheduler survive? Here we consider systems with $N \geq n$ words (programs) and any $n$ out of the $N$ should constitute a solution of MC. We call these $MC(n, m)$ systems with $N$ words.

   **Theorem 3** *For every $N \geq n$, almost every choice of $N$ random words of length $cn \log N$ in an alphabet of $m = 7n$ letters is an $MC(n, m)$ system with $N$ full words (words that contain every letter in $1, \ldots, m$). Moreover, every schedule on these words terminates in $O(n \log N)$ steps. Here c is an absolute constant.*

7. Since we are dealing with full words (words that contain every letter in $1, \ldots, m$) and we seek to make them short, we are ultimately led to consider the case where each word is a permutation on $[m]$. At the moment the main reason to study this question is its aesthetic appeal. We can design permutation-based oblivious $MC(n, 2n-1)$ algorithms for very small $n$ (provably for $n = 3$, computer assisted proof for $n = 4$). We suspect that no such constructions are possible for large values of $n$, but we are unable at present to show this. We do know, though that **Theorem 4** *For every integer $d \geq 1$ there is a collection of $N = n^d$ permutations on $m = cn$ symbols such that every $n$ of these permutations constitute an oblivious $MC(n, m)$ algorithm. The constant c depends only on d. In fact, this holds for almost every choice of $N$ random permutations on $[m]$.*

We should stress that our proofs of Theorems 3 and 4 are purely existential. The explicit construction of such systems of words remains largely open, though we do have some results in this direction, e.g.,

**Theorem 5** *For every integer $d \geq 1$ there is an explicitly constructed collection of $N = n^d$ permutations on $m = O_d(n^2)$ symbols such that every $n$ of these permutations constitute an oblivious $MC(n, m)$ algorithm.*

## 1.1 A road map

Most of the technical results in this paper concern the design of oblivious algorithms for the MC task, either with the least possible number of chairs (namely, $m = 2n - 1$), or only $m = O(n)$ chairs. These results then extend to the renaming task. The purpose of this section is to highlight several additional aspects of the subject.

We start with a number of simple observations. (i) An oblivious $MC(n, m)$ algorithm cannot include any two identical words. Otherwise the corresponding players might move together in lock-step, constantly being in collision. Hence it is essential that no two processors have the same program. (ii) For every oblivious $MC(n, m)$ algorithm with finite words, there is a finite upper bound on the number of moves a processor can make before termination. This is because there are only finitely many system states, and in a terminating sequence of moves no system state can be visited twice. (iii) In fact, for every collection of finite words there is a directed graph whose vertices are all the system states. Edges correspond to the possible transitions. The collection of words constitute an oblivious MC protocol iff this graph is acyclic. These observations depend on the assumption that the algorithm is deterministic.

Our oblivious algorithms for MC have a number of additional desirable properties. For every $n > 1$, $m = 2n - 1$ and $N > n$ we design $N$ periodic words that are full (i.e., contain every chair) with the following properties: for every choice of $n$ or fewer of the $N$ words, for every choice of states on these words, each word is guaranteed to reach a chair not shared by any other word. There is an upper bound (that depends only on $N$) on the number of steps taken by any word, and moreover, this guarantee holds even if other words fault and no longer change states. Hence our oblivious algorithms can be run in dynamic settings in which the set of players in the system keeps changing. It is still guaranteed to reach a conflict free state provided that there are sufficiently long intervals without dynamic changes. Moreover, this protocol can withstand various kinds of faults, e.g., non-faulty processors can complete their computations even in the presence of faulty processors. To illustrate this idea, consider a company that manufactures $N$ communication devices, each of which can use any one of $m$ frequencies. If several such devices happen to be at the same vicinity, and simultaneously transmit at the same frequency, then interference occurs. Devices can (i) Move in or out of the area, (ii) Hop to a frequency of choice and transmit at this frequency, (iii) Sense whether there are other transmissions in this frequency. The company wants to provide the following guarantee: If no more than $n$ devices reside in the same geographical area, then no device will suffer more than $M$ interference events for some specific integer $M$. Our oblivious MC algorithms would guarantee this by pre-installing in each device a list of frequencies (a *word* in our terminology), and having the device hop to the next frequency on its list (in a cyclic fashion) in response to any interference it encounters. No communication beyond the ability to sense interference is needed.

In Section 2 we present a formal model in which our oblivious algorithms apply, placing it within a known standard framework for distributed computing. The model presented in Section 2 does not attempt to capture all possible interpretations of our MC protocols. For example, the model concerns tasks that terminate, whereas our protocols work equally well in reactive systems that keep adapting to a changing environment. What the formal model does capture is important connections with previous works in distributed computing, from which a lower bound of $m \geq 2n-1$ can easily be inferred. This lower bound shows that our upper bounds are best possible, so let us elaborate on it.

Not all aspects of oblivious protocols are required for the purpose of the lower bound $m \geq 2n-1$. The two crucial aspects are the asynchrony of the model, and the fact that our algorithms are deterministic (no randomization). In a synchronous setting, where in every time step, every processor involved in a collision moves to its next state), $m = n$ suffices, even for oblivious protocols. (This can be proven using the techniques of Theorem 3. Details omitted.) Likewise, $m = n$ suffice if randomization is allowed – with probability 1 eventually there are no collisions. However, no specific upper bound on the number of steps can be guaranteed in this case. Moreover, if the randomized algorithms is run using pseudorandom generators (rather than true randomness) the argument breaks. For any fixed seed of a pseudorandom generator, the algorithm becomes deterministic and the lower bound $m \geq 2n - 1$ holds.

The lower bound of $m \geq 2n - 1$ uses some benign-looking aspects of the MC task, so further discussion is called for. Recall that each processor starts in an arbitrary chair, dictated by the input. In the absence of an external input specifying the starting chair, a trivial oblivious MC algorithm (with $m = n$) contains $n$ distinct single-letter words. Another requirement is that if the input chairs are all unique, all processors are good and the input is the output. Without such a requirement, the processors might simply ignore the initial input and the trivial oblivious MC algorithm would still apply. Hence the lower bound of $m \geq 2n-1$ depends on requirements beyond the need for each processor to capture a chair in exclusion. Here this extra requirement is the possibility to dictate an input. This particular requirement is common in distributed computing as it allows composition of protocols. It also makes it easy to transfer previously existing lower bounds to our MC problem.

Our present proof for the lower bound of $m \geq 2n-1$ leaves something to be desired. It relies on previous nontrivial work in distributed computing. What's worse is that we prove a lower bound for a simple oblivious model via a reduction to a lower bound proved in a more complicated model. This roundabout approach obscures the essential properties that make the lower bound work. Indeed, in a companion manuscript (in preparation), we present a self contained proof for the lower bound of $m \geq 2n - 1$. That presentation clarifies the minimal requirements that are needed in order to make the lower bound work. In particular, it is not necessary that one can dictate an arbitrary starting chair for each processor – dictating one of two chairs suffices.

As noted, we design oblivious $MC(n, m)$ protocols with $m = 2n - 1$. Part of our work also concerns analyzing what ratios between $m$ and $n$ one can obtain using collections of randomly chosen words as in Theorem 3. As explained in the introduction, this allows us to present more efficient deterministic oblivious programs – though random words seem to need more chairs, they can reach conflict free configurations more quickly. Moreover, the use of random words is a design principle that can be applied to design oblivious algorithms for other tasks as well. Developing an understanding of what they can achieve and techniques for their analysis is likely to pay off in the long run. One of the major questions that remain open in our work is whether randomly chosen words can be used to design deterministic oblivious MC protocols with $m = 2n - 1$.

## 2 The model

### 2.1 Tasks

A task [13] is a distributed computational problem involving several processes (or processors). There is an upper bound denoted by $n$ on the number of processes that may participate in the task. Each participating process starts with a private input value, exchanges information with other participating processes (for example, by writing to and reading from a common memory), and halts with an output value. A nonparticipating process is indistinguishable to other processes from a process that is participating but has not yet performed any observable operation (such as a write operation). The task is specified by a relation $\Delta$ that associates with every input vector (one element per participating processor) a set of output vectors that are allowed given this input. For notational convenience, the input and output vectors are of dimension $n$ (even when the number of participating processors is smaller) and the corresponding entries for nonparticipating processors are denoted there by the special symbol $\bot$. We use the notation $v_{inp}$ and $v_{out}$ to denote these vectors, though the reader should note that the subscripts $inp$ and $out$ might be a bit misleading (the $\bot$ entries for nonparticipating processors are neither true inputs nor true outputs, but only notation indicating that the processors are not participating). Given our convention regarding $\bot$, an input vector $v_{inp}$ implicitly describes which are the participating processors, namely, $Prtc(v_{inp}) = \{p_i \mid v_{inp}(i) \neq \bot\}$. Restating our conventions regarding notation for nonparticipating processors we have that for $(v_{inp}, v_{out}) \in \Delta$ it must hold that $v_{inp}(i) = \bot$ iff $v_{out}(i) = \bot$.

**The Musical Chairs task:** In the musical chairs task $MC(n, m)$ there are $n \geq 1$ processors $\{p_1, \ldots, p_n\}$, and a set of *chairs* $\{1, \ldots, m\}$. Each participating processor starts in an arbitrary chair, dictated by its input, and it has to capture a chair in exclusion. If the input chairs are all unique, all processors must output their input. The formal definition, following the notations of [13] is:
$v_{inp}(i), v_{out}(i) \in \{1, 2, ..., m, \bot\}$.

1. If $\forall i, j \in Prtc(v_{inp}), v_{inp}(j) \neq v_{inp}(i)$ then $v_{out} = v_{inp}$, and
2. Else $\forall i, j \in Prtc(v_{inp}), v_{out}(j) \neq v_{out}(i)$.

**The Renaming task:** In the $Renaming(n, m)$ task there are $n \geq 1$ processors $\{p_1, \ldots, p_n\}$, and $m$ slots numbered $1, \ldots, m$. Each participant has to capture a slot in exclusion. Formally, the processors have no input, though for notational convenience we shall assume that participating processors have the input 1. If only $k < n$ processors participate, then each has to capture (output) a unique slot from the first $\min(2k - 1, m)$ slots. If all the $n$ processors participate then they each capture one of the $m$ slots in exclusion. The formal definition, following the notations of [13] is:
$v_{inp}(i) \in \{1, \bot\}$ and $v_{out}(i) \in \{1, 2, ..., m, \bot\}$

1. If $|Prtc(v_{inp})| = k < n$ then $v_{out}(i) \in \{1, 2, ..., 2k - 1, \bot\}$ and $\forall i, j \in Prtc(v_{inp}), v_{out}(j) \neq v_{out}(i)$, and
2. If $|Prtc(v_{inp})| = n$ then $v_{out}(i) \in \{1, 2, ..., m, \bot\}$ and $\forall i, j \in Prtc(v_{inp}), v_{out}(j) \neq v_{out}(i)$.

## 2.2 The Oblivious Model

The *Oblivious* model is an asynchronous distributed computing model in which each processor, at each point of time, exposes an output value it currently proposes, and may receive at most one bit of information. This bit indicates whether its proposed output is legal with respect to the other currently proposed outputs (and hence the processor may halt) or not (and then the processor should continue the computation). If a processor decides to halt at the current state, then its proposed output is its final output. We denote the set of possible output values by $O$. A *system configuration* (or configuration for short) is a vector of $n$ elements, one per processor, whose entries come from the set $O \cup \{\bot\}$. Here $\bot$ represents a processor that has not yet proposed any output, either because it is not participating, or because it was not scheduled yet to propose an output (these two cases are indistinguishable to other processors. An entry from $O$ represents the output a corresponding processor proposes in the configuration. In an oblivious algorithm correctly designed for a given task, eventually all participating processors must halt, and the final configuration must be a legal output vector in the task specification.

The defining feature of the *oblivious* model is that each processor may receive only one bit of information about the system configuration in each computation step; whether the current configuration is illegal and it should change its state (and thus its proposed output), or whether it may halt in its current state. The fact that a processor $p_i$ is not informed to change its state does not necessarily mean that the current configuration is legal. For example, the configuration might be illegal, but changing $p_i$'s state would not get the system any closer to a legal configuration. However, in a correct algorithm (program) at least one processor is notified to change its state in an illegal configuration. The choice of function specifying for each configuration which processors may change their state and which may halt and output is up to the algorithm designer. The algorithm provides each processor with a predicate on configurations, specifying in which configuration it changes its state, and in which it may halt. In the most general setting the predicate provided for each processor may depend on its input, possibly a different predicate for different inputs. However, throughout an execution one predicate is used for each processor. Our formal model does not exclude the use of arbitrary complex predicates, but oblivious algorithms have greater appeal when the predicates involved are simple and natural. For the two tasks considered in this paper, the same collision predicate is used by all the processors for any input.

Initially, and as a function of its input, each processor $p_i$ selects a word $\pi_i$ over $O$, and a predicate $pred_i$ on the set of of all configurations. The first letter in $\pi_i$ is $p_i$'s input, i.e., $\pi_i[1] = input_i \in O$. For tasks such as renaming in which a processor need not have any input, the first letter is set to be an output that is valid if no other processor participates (which explains why in the definition of renaming we used the convention that the input to participating processors is 1).

We describe the system using the notion of an omnipotent know-all scheduler called *asynchronous* (other schedulers with different names are described in the sequel). Execution under the control of the asynchronous scheduler proceeds in rounds. The scheduler maintains a set $P$ of participating processors, a set $E \subset P$ of enabled processors, and a set $DONE$ (disjoint from $P$) of processors that have already halted. These sets are initially empty. In each round the scheduler performs the following sequence of operations. It may add some not yet participating processors to $P$. It may evaluate the predicate $pred_i$ for some subset of processors in $P \setminus E$. If $pred_i$ evaluates to true, the scheduler adds processor $p_i$ to the set $E$. Otherwise, if it evaluates to false, it removes $p_i$ from $P$ and adds it to the set $DONE$. Finally, the scheduler selects a subset $SE \subseteq E$, removes it from $E$, and moves each $p_i \in SE$ to its next letter in $\pi_i$. I.e., the current output of $p_i$ is replaced by the next one in its program, $\pi_i$. This completes the round.

An oblivious algorithm solves a task if for every input vector, the scheduler is forced to eventually place all participating processors in the $DONE$ set. At that point it can no longer continue, and the final configuration is such that $(v_{inp}, v_{out}) \in \Delta$, the relation that defines the task.

A well known model for distributed computing is the read/write wait-free model, that we shall sometimes simply refer to as read/write. The main features of this model is that processors communicate via read and write operations, scheduling is asynchronous, and every task is completed by a processor in a finite number of steps (regardless of the actions of other processors; this is the wait-free property). See [12] or [13] for more details. The asynchronous scheduler for oblivious algorithms mimics the behavior of an asynchronous read/write algorithm on configurations. Thus Theorem 6 below can be proved simply by having each processor emulate the scheduler through reads (snapshots) and writes of its newly proposed output in shared memory.

**Theorem 6** *Every task that is solvable obliviously is solvable read-write wait-free.*

**Proof.** Given an oblivious distributed program to solve a task we provide a read-write wait-free algorithm to solve the same task. In the read-write system the shared memory has one single writer multi reader register for each processor, in which the processor publishes its currently proposed output. W.l.o.g., we can replace each read by an atomic snapshot [3].

Initially, as a function of its input, each processor writes its first proposed output, and uploads its predicate for the run. Then the processor repeatedly takes a snapshot and writes its next output in its oblivious program until a snapshot evaluates to false. A snapshot evaluated by the predicate to false, corresponds to a configuration in which a processor was added to the $DONE$ set.

An execution in the read-write model is thus a linear sequence of reads (snapshots) and writes and it corresponds to an execution in the oblivious model in the following way: All the processors that observe the same snapshot are those that the asynchronous scheduler evaluates their predicate at the same round. Those evaluated to true are added to the enabled set $E$, and those evaluated to false are added to the $DONE$ set and stopped forever. The set of writes that occur after this snapshot, and before the next snapshot, correspond to the subset of enabled processors that the scheduler move to their next letter in their program $\pi$. Since the scheduler must stop with the correct output vector so will the read/write algorithm. ∎

Thus the oblivious model is subsumed by the read/write model. Is this a proper inclusion? To clarify the answer we introduce an intermediate class of tasks that we call Output Negotiation, or $ON$. It includes those tasks solvable read-write wait-free in a system where writing is in the oblivious model (processors can only expose their proposed outputs), whereas reading is as in the general read/write model (a processor can read all exposed information rather than only a single predicate). By definition, every obliviously solvable task is $ON$ solvable.

**Corollary 7** *Every obliviously solvable task is in $ON$.*

Obviously, $ON$ is a subset of read/write, and in Theorem 8 below we show that this inclusion is proper. Consequently the oblivious model is a strict subset of read/write.

**Theorem 8** *There exists a task, AntiMC, that is solvable read-write wait-free but does not belong to $ON$.*

**Proof.** The task *AntiMC* is a variation on epsilon agreement [7]. It is a task with 3 processors whose input and output are each a number in $\{1, \ldots, 5\}$. A processor running alone must output its input. Otherwise all the outputs must be one of two consecutive numbers (5 and 1 are *not* consecutive). Formally, AntiMC on 3 proccesors:
$v_{inp}(i), v_{out}(i) \in \{1, 2, ..., 5, \bot\}$

1. If $|Prtc(v_{inp})| = 1$ then, $v_{inp} = v_{out}$,

2. Else $\forall i, j \in Prtc(v_{inp}), |v_{out}(j) - v_{out}(i)| \leq 1$.

AntiMC is solvable read-write wait-free using the standard approach for solving $\epsilon$-agreement. Let us provide a few more details. Participating processors do not only post a proposed output, but also an integer *weight* (in the range 1 to $W$) that specifies how "confident" they are in their output. We now describe the actions of a participating processor. Initially, it posts its input as a proposed output, and posts a weight of 1. Thereafter, the processor performs "rounds" in its own speed (determined by an asynchronous scheduler). In a round the processor inspects the proposed outputs and posted weights of all other processors (assigning weight 0 to those processors who did not yet post anything), computes a weighted average of all proposed outputs (including its own), and posts the integer nearest to it as a new proposed output. It also raises its weight by 1 and posts its new weight. Processors halt (with their current proposed output as their final output) when the total weight reaches $W$. Choosing $W$ to be sufficiently large guarantees that all final outputs are within 1 of each other. Further details are omitted.

We now show that AntiMC is not solvable just by communicating outputs. Observe that we may assume that a processor first posts its input. (If a processor performs read operations before posting any output we may schedule the read operations before any other processor posted an output, and hence eventually the processor is forced to post its input.) Consider the input vector $(1, 5, 3)$ and two scheduling scenarios. In the first scenario, schedule $p_1$ first (with input 1) and continue to schedule only $p_1$. Eventually $p_1$ must terminate at 1. Now schedule $p_2$ (with input 5) and let it post its input. In the second scenario reverse the roles of $p_2$ to terminate with 5 and $p_1$ to have just posted 1. Observe that in the first scenario the outputs should eventually be in $\{1, 2\}$ and in the second scenario in $\{4, 5\}$. Now schedule $p_3$ (with input 3) and let it run without interference until termination. Both scenarios are indistinguishable to $p_3$, and whatever it outputs is incompatible with at least one of the scenarios. ∎

## 2.3 Impossibility of $MC(n, 2n-2)$

In Sections 3 and 4 we show that $MC(n, 2n-1)$ and Renaming$(n, 2n-1)$ are solvable obliviously. Renaming$(n, 2n-2)$ is unsolvable read-write wait-free [8, 11], and hence not solvable obliviously either. Theorem 9 shows a reduction from Renaming$(n, 2n-2)$ to $MC(n, 2n-2)$. This implies that $MC(n, 2n-2)$ is not solvable read-write wait-free, and hence also not solvable obliviously.

**Theorem 9** *Renaming*$(n, 2n-2)$ *is read-write wait-free reducible to* $MC(n, 2n-2)$.

**Proof.** Whenever we say *algorithm* in the proof, we shall mean a read/write wait-free distributed algorithm.

Suppose that there is an algorithm for the $MC(n, 2n-2)$ task. Recall that there is an algorithm for the Renaming$(n-1, 2n-3)$ task. By using both algorithms, we shall design an algorithm for the Renaming$(n, 2n-2)$ task. The basic observation is that if fewer than $n$ processors participate then Renaming$(n, 2n-2)$ is equivalent to Renaming$(n-1, 2n-3)$, and if $n$ processors participate Renaming$(n, 2n-2)$ is equivalent to $MC(n, 2n-2)$. This suggests incorporating a "counting" task that helps processors determine how many processors are participating, and based on the outcome of the counting task they decide whether to participate in Renaming($n-1, 2n-3$) or in $MC(n, 2n-2)$.

We now provide more details. Each task is run independently (e.g., on different portions of shared memory) so that there is no interference among tasks. When a processor first arrives (meaning that it participates in Renaming($n, 2n-2$)) it performs the counting task. In this task it first announces its arrival (e.g., by writing its ID into some specific location in shared memory). Thereafter it counts (by reading the corresponding locations in shared memory) how many other processors have arrived. If the count shows that the total number of arriving processors (including itself) is $n$, the processor joins the $MC(n, 2n-2)$ task, with a private input value of 1. If on the other hand the count is smaller, the processor first joins the Renaming($n-1, 2n-3$) task (again with input 1). However, a processor that completes the Renaming($n-1, 2n-3$) task is not done (because by the time of its arrival and the time that it completed the Renaming($n-1, 2n-3$) task it could be that additional processors arrived and are running the $MC(n, 2n-2)$ task, thus leading to incompatibilities in the outputs). Instead, it joins the $MC(n, 2n-2)$ task, using his output from Renaming($n-1, 2n-3$) as input to $MC(n, 2n-2)$.

It is not difficult to verify the following claims:

1. The total number of processors that ever run the Renaming($n-1, 2n-3$) task is at most $n-1$, and hence this task runs properly.

2. If the total number of participating processors is at most $n-1$ then the output is that of the Renaming($n-1, 2n-3$) task, and hence legal for Renaming($n, 2n-2$).

3. If the total number of participating processors is $n$ then the output is that of the $MC(n, 2n-2)$ task, and hence legal for Renaming($n, 2n-2$).

The above claims imply the theorem. ∎

## 2.4 Cyclic Finite Program (Word)

The definition of oblivious algorithms in Section 2.2 postulates that as a function of its input, each processor selects an (infinite) sequence of outputs. For the Renaming task, processors have no input (or alternatively, are assumed to always have the input 1), and hence each processor has only one sequence. For $MC$ there are $m$ possible inputs that a processor may have, and hence our model allows each processor to have $m$ different sequences, one for each input. Nevertheless, our constructions of oblivious algorithms all have the property that the same sequence is used for all inputs. Moreover, we consider finite sequences over which the processor goes cyclically. In the $MC$ task one can designate $m$ locations in the word, each corresponding to a possible output that has been dictated by the input to the processor. The infinite word for each input is then attained by advancing cyclically on the word starting from that designated location. In fact, we strengthen the scheduler; If an output appears in the word more than once, every appearance of the output is a valid starting point for the $MC$ program (providing the scheduler with more choices). This makes the $MC$ program self-stabilizing [5, 6], as mentioned in the introduction.

## 2.5 Simplified Oblivious Model for MC and Renaming

Our general model for oblivious algorithms is described using the *asynchronous* scheduler (Section 2.2). The asynchronous scheduler enjoys a large degree of freedom in choosing which processor to move. To simplify the design and analysis of oblivious algorithms, it is convenient to consider simpler schedulers that have fewer degrees of freedom, but are nevertheless equivalent to the asynchronous scheduler in their power to prevent successful completion of tasks. Our oblivious algorithms for MC and Renaming use only a simple collision predicate. That is, a processor can become enabled by the asynchronous scheduler only if it is involved in a collision, and may be moved to the $DONE$ set only if not involved in a collision. The simple nature of this collision predicate allows us to present a sequence of schedulers that appear to be successively weaker, though all are in fact equivalent (with respect to MC and Renaming). The results in this section will be presented only with respect to the MC task, but at the end of this section we explain how to extend them to Renaming.

**Terminology.** Whenever we say that two schedulers are *equivalent* it means that a collection of $n$ words over an alphabet of $m$ chairs forms an oblivious $MC(n,m)$ algorithm with respect to one scheduler if and only if it forms an oblivious $MC(n,m)$ algorithm with respect to the other scheduler. I.e., one scheduler has an infinite run from some initial configuration with a set of $n$ words if and only if the other scheduler has.

Recall that the asynchronous scheduler maintains several sets, $P$ (for those processors that are participating), $E$ (for those processors that may move at the current or some future round), and $DONE$ (for those processors that will move no more). The set $E$ gives the asynchronous scheduler its flexibility and freedom to move processors that have been in conflict at some point in the future. We now present a scheduler that makes only limited use of the set $P$, and does not use the set $DONE$.

*Quiescent scheduler.* It is a scheduler for which the set $P$ never changes. That is, every processor that participates in the execution is added to $P$ (and posts a proposed output) immediately as the execution begins (rather than at a point in time determined by the scheduler). Moreover, no processor is ever told to halt (and hence there is no need for the set $DONE$). Processors that are never told to move from some point, simply become quiescent. When all processors are quiescent the scheduler has no more moves, and the execution terminates. Other than putting all participating processors in $P$ upfront and not having a $DONE$ set, the quiescent scheduler behaves like the asynchronous scheduler. An oblivious $MC(n,m)$ protocol is required to force the quiescent scheduler to reach a configuration in which $E$ is empty and there are no collisions. Conversely, a quiescent scheduler foils a proposed oblivious algorithm if it can generate an infinite execution in which in every configuration either $E$ is nonempty or there is a collision.

**Proposition 10** *The asynchronous scheduler and the quiescent scheduler are equivalent.*

**Proof.** *Asynchronous scheduler at least as strong as quiescent scheduler.* All moves available to the quiescent scheduler are also available to the asynchronous scheduler. Hence if the quiescent scheduler has an infinite run, the asynchronous scheduler can force an infinite execution as well (by imitating the quiescent scheduler).

*Quiescent scheduler at least as strong as asynchronous scheduler.* For an asynchronous scheduler to foil a proposed oblivious $MC(n,m)$ algorithm, it needs to generate an infinite execution. The quiescent scheduler can imitate the asynchronous scheduler with the following differences. Whenever the asynchronous scheduler places a processor in the $DONE$ set (there are at most $n$ rounds in which this happens), the quiescent scheduler does not do so (and drops the round if no other action was taken in this round). Whenever the asynchronous scheduler places a processor in $P$ (there are at most $n$ rounds in which this happens), the quiescent scheduler instead places the processor in $P$ in the first round. All other moves of the asynchronous scheduler remain legal for the quiescent scheduler, and hence infinite executions for the asynchronous scheduler result in infinite executions for the quiescent scheduler. ∎

Our next goal is to get rid of the set $E$.

*Immediate scheduler.* The immediate scheduler is similar to the quiescent scheduler, except that it does not maintain a set $E$ of enabled processors. Instead, in each round it can only select processors that are currently involved in a collision and move them. It is important to note that in a round the immediate scheduler does not need to select all processors that are involved in a collision – it may select a nonempty subset of its choice. An oblivious $MC(n,m)$ protocol is required to force the immediate scheduler to reach a configuration in which there are no collisions. Conversely, an immediate scheduler foils a proposed oblivious $MC(n,m)$ algorithm if it can generate an infinite execution never reaching a configuration in which there are no more collisions.

**Proposition 11** *The quiescent scheduler and the immediate scheduler are equivalent.*

**Proof.** The immediate scheduler is a special case of the quiescent scheduler (essentially it places processors in $E$ and moves them at the same round). Hence it remains to show that the immediate scheduler is at least as strong as the quiescent scheduler. This is equivalent to showing the following statement: whenever there is an infinite run of the quiescent scheduler, there is also an infinite run of a quiescent scheduler in which whenever it places a processor in $E$, it moves it in the same round. We prove this last statement by a double induction on the round $t$ (increasing) and the number $k$ of processors that violate this statement in round $t$ (decreasing until $k = 0$, and thus causing $t$ to increase). Our inductive proof has the property that some rounds might become empty in the process (contain no action on behalf of the scheduler). However, despite this, every infinite run transforms into an infinite run, because the number of processor moves is kept unchanged.

Given a proposed oblivious $MC(n,m)$ algorithm and an infinite execution by the quiescent scheduler, let $t$ be the first round in which there is a processer added to $E$ and not moved in the same round, and let $k \geq 1$ be the number of such processors in round $t$. Pick an arbitrary processor $p$ added to $E$ in round $t$ and not moved in this round. If $p$ is not moved even in any future round, simply do not put $p$ in $E$. This decreases $k$ and the inductive step is done. Alternatively, if $p$ is moved in a future round, say round $t' > t$, we consider two cases. In one case there is some round $t"$ with $t < t" \leq t'$ in which $p$ is involved in a collision. In this case, rather than placing $p$ in $E$ in round $t$, simply do this in round $t" > t$ instead. This decreases $k$ and the inductive step is done. In the other case, there is no such round $t"$. In this case, move $p$ in round $t$ rather than round $t'$. This also decreases $k$ by one. Observe that all moves available to the scheduler between rounds $t$ and $t'$ are still available also after this change in the scheduler, since $p$ could not contribute to enabling processors within this interval of rounds. ∎

Having eliminated the sets $E$ and $DONE$, we now turn our attention to limiting the number of processors that can be moved in a round.

*Pairwise immediate scheduler.* This is similar to the immediate scheduler but with the following restriction. In every round, the pairwise immediate scheduler can select any two processors currently in collision with each other, and move either one of them, or the other, or both. Equivalently, in every round either only one processor (involved in a collision) moves, or two processors that share the same chair.

**Proposition 12** *The immediate scheduler and the pairwise immediate scheduler are equivalent.*

**Proof.** The pairwise immediate scheduler is a special case of the immediate scheduler. Hence it remains to show that whenever there is an infinite run with the immediate scheduler, there is also an infinite run with the pairwise immediate scheduler. We prove this last statement by a double induction on the round $t$ and the number $k$ of processors that move in round $t$.

Given a proposed oblivious $MC(n,m)$ algorithm and an infinite execution by the immediate scheduler, let $t$ be the first round in which the moves were not consistent with a pairwise immediate scheduler and let $k$ be the number of processors that move in round $t$. There are two cases to consider. In one case the set SE of processors that moved shared in round $t$ the same chair and $k \geq 3$. Break round $t$ into two rounds, pushing future rounds by one. In the first of them (round $t$) move only one of the processors from SE and in the second round (round $t+1$) move the rest (they can still move because there are at least two of them). This completes the inductive step with respect to $t$. The other case is that the set SE of processors that moved in round $t$ collided on at least two different chairs. Pick one of these chairs, say chair $c$, and let $SE(c)$ be the set of those processors in $SE$ that in round $t$ collide in chair $c$. Break round $t$ into two rounds, pushing future rounds by one. In the first of them (round $t$) move only those processors in $SE(c)$, and in the second round (round $t+1$) move those processors in $SE - SE(c)$. This completes the inductive step (as either $k$ decreased or $t$ increased). ■

The use of the pairwise immediate scheduler (which as we showed is equivalent to the asynchronous scheduler) helps simplify the proofs of theorems 1, 2 and 4. However, for the proof of Theorem 3 even the pairwise immediate scheduler has too many degrees of freedom. It is true that it has to pick only one pair of processors to move (and then either move only one or both of them), but it is still free to pick a pair of its choice (among those pairs that collide). We would like to eliminate this degree of freedom.

*Canonical Scheduler.* The canonical scheduler is similar to the pairwise immediate scheduler but with the following difference. In every round in which there is a collision, one designates a *canonical pair*. This is a pair of processors currently in collision with each other, but they are not chosen by the scheduler, but rather dictated to the scheduler. Given the canonical pair, the scheduler can move either one of the these processors, or the other, or both. But how is the canonical pair chosen? In the current paper this does not really matter to us, as long as the choice is deterministic. For concreteness, we shall assume the following procedure. Consider all pairs of processors and fix an arbitrary order on them. In a configuration with a collision, the *canonical pair* is the first pair of players in the order that share a chair.

We now prove the equivalence of the canonical scheduler with the immediate scheduler (the proof does not become any simpler if we replace in it immediate scheduler by pairwise immediate scheduler).

**Proposition 13** *The immediate scheduler and the canonical scheduler are equivalent.*

**Proof.** The canonical scheduler is a special case of the immediate scheduler. Hence it remains to show that whenever there is an infinite run of the immediate scheduler, there is also an infinite run with the canonical scheduler. We prove this last statement by induction on the round $t$.

Given a proposed oblivious $MC(n,m)$ algorithm and an infinite execution by the immediate scheduler, let $t$ be the first round in which the moves were not consistent with a canonical scheduler. That is, the canonical pair at round $t$ consists of two processors (say $P_1$ and $P_2$, without loss of generality) that collide on a chair (say, chair $c_1$), whereas the immediate scheduler moved at least one processer not from the canonical pair. We consider several cases.

*Case 1.* The immediate scheduler never moves $P_1$ in any round from $t$ onwards. In this case move $P_2$ in round $t$. Note that all moves (except for the move just performed, moving $P_2$ away from $c_1$) performed by the immediate scheduler from round $t$ onwards are still available to this scheduler (because chair $c_1$ remains occupied). Hence the total number of moves in the schedule did not change, whereas $t$ increases by one, completing the inductive step. The same argument can be applied with $P_1$ and $P_2$ exchanged.

*Case 2.* The immediate scheduler moves $P_2$ out of $c_1$ in a later round than it moves $P_1$. In this case move $P_1$ in round $t$. Again, all moves (except for the move just performed, moving $P_1$ away from $c_1$) performed by the immediate scheduler from round $t$ onwards are still available to this scheduler. The same argument can be applied with $P_1$ and $P_2$ exchanged.

*Case 3.* The immediate scheduler moves both $P_1$ and $P_2$ out of $c_1$ in the same round $t' \geq t$. There are two subcases to consider. In one, there is no processor other than $P_1$ and $P_2$ on chair $c_1$ in any of the rounds $t, \ldots, t'$. In this subcase, move $P_1$ and $P_2$ in round $t$ (pushing future rounds by one). All moves performed by the immediate scheduler from round $t$ to $t'$ are still available to this scheduler. The other subcase is that there is some round $t \leq t" \leq t'$ in which some other processor say $P_3$ is on chair $c_1$. Consider the largest such $t"$. Move $P_1$ in round $t$ (pushing future rounds by one) and $P_2$ in round $t" + 1$ (the round that previous to the pushing of rounds was round $t"$), together with whoever else is moved at that round. ∎

**Remark.** The results of this section apply also for Renaming and not only for $MC$. However, the proofs for Renaming need to be slightly changed. The difference is that in Renaming an oblivious algorithm fails not only if the scheduler manages to exhibit an infinite execution, but also if the scheduler manages to make a processor output a value larger than $2k - 1$ when $k$ is the number of participating processors. Modifying the proofs so that they handle also this form of failure is straightforward, and we omit the details.

## 3 An oblivious MC algorithm with $2n - 1$ chairs

### 3.1 Preliminaries

In this section we prove the upper bound that is stated in Theorem 1. We start with some preliminaries. The length of a word $w$ is denoted by $|w|$. The concatenation of words is denoted by $\circ$. The $r$-th power of $w$ is denoted by $w^r = w \circ w \ldots \circ w$ ($r$ times). Given a word $\pi$ and a letter $c$, we denote by $c \otimes \pi$ the word in which the letters are alternately $c$ and a letter from $\pi$ in consecutive order. For example if $\pi = 2343$ and $c = 1$ then $c \otimes \pi = 12131413$. A collection of words $\pi_1, \pi_2, \ldots, \pi_n$ is called *terminal* if no schedule can fully traverse even one of the $\pi_i$. Note that we can construct a terminal collection from any $MC$ algorithm just by raising each word to a high enough power.

We now introduce some of our basic machinery in this area. We first show how to extend terminal sets of words.

**Proposition 14** *Let $n, m, N$ be integers with $1 < n < m$. Let $\Pi = \{\pi_1, \ldots \pi_N\}$ be a collection of $m$-full words such that*

$$\text{every } n \text{ of these words form an oblivious } MC(n,m) \text{ algorithm.} \tag{1}$$

*Then $\Pi$ can be extended to a set of $N + 1$ $m$-full words that satisfy condition (1).*

**Proof.** Suppose that for every choice of $n$ words from $\Pi$ and for every initial configuration no schedule lasts more than $t$ steps. (By the pigeonhole principle $t \leq L^n$, where $L$ is the length of the longest word in $\Pi$). For a word $\pi$, let $\pi'$ be defined as follows: If $|\pi| \geq t$, then $\pi' = \pi$. Otherwise it consists of the first $t$ letters in $\pi^r$ where $r > |\pi|/t$. The new word that we introduce is $\pi_{N+1} = \pi_1' \circ \pi_2' \circ \ldots \circ \pi_n'$. It is a full word, since it contains the full word $\pi_1$ as a sub-word.

We need to show that every set $\Pi'$ of $n-1$ words from $\Pi$ together with $\pi_{N+1}$ constitute an oblivious $MC(n,m)$ algorithm. Observe that in any infinite schedule involving these words, the word $\pi_{N+1}$ must move infinitely often. Otherwise, if it remains on a letter $c$ from some point on, replace the word $\pi_{N+1}$ by an arbitrary word from $\Pi - \Pi'$ and stay put on the letter $c$ in this word. This contradicts our assumption concerning $\Pi$. (Note that this word contains the letter $c$ by our fullness assumption.) But $\pi_{N+1}$ moves infinitely often, and it is a concatenation of $n$ words whereas $\Pi'$ contains only $n-1$ words. Therefore eventually $\pi_{N+1}$ must reach the beginning of a word $\pi_\alpha$ for some $\pi_\alpha \notin \Pi'$. From this point onward, $\pi_{N+1}$ cannot proceed for $t$ additional steps, contrary to our assumption. ∎

Note that by repeated application of Proposition 14, we can construct an arbitrarily large collection of $m$-full words that satisfy condition (1).

We next deal with the following situation: Suppose that $\pi_1, \pi_2, ..., \pi_m$ is a terminal collection, and we concatenate an arbitrary word $\sigma$ to one of the words $\pi_i$. We show that by raising all words to a high enough power we again have a terminal collection in our hands.

**Lemma 15** *Let $\pi_1, \pi_2, ..., \pi_p$ be a terminal collection of full words over some alphabet. Let $\sigma$ be an arbitrary full word over the same alphabet. Then the collection*

$$(\pi_1)^k, (\pi_2)^k, ..., (\pi_{i-1})^k, (\pi_i \circ \sigma)^2, (\pi_{i+1})^k, ..., (\pi_p)^k$$

*is terminal as well, for every $1 \leq i \leq p$, and every $k \geq |\pi_i| + |\sigma|$.*

**Proof.** We split the run of any schedule on these words into *periods* through which we do not move along the word $(\pi_i \circ \sigma)^2$. We claim that throughout a single period we do not traverse a full copy of $\pi_j$ in our progress along the word $(\pi_j)^k$. The argument is the same as in the proof of Proposition 14. By pasting all these periods together, we conclude that during a time interval in which we advance $\leq |\pi_i| + |\sigma| - 1$ positions along the word $(\pi_i \circ \sigma)^2$ every other word $(\pi_j)^k$ traverses at most $|\pi_i| + |\sigma| - 1$ copies of $\pi_j$. In particular, there is a whole $\pi_j$ in the $j$-th word in the collection that is never visited. If the schedule ends in this way, no word is fully traversed, and our claim holds.

So let us consider what happens when a schedule makes $\geq |\pi_i| + |\sigma|$ steps along the word $(\pi_i \circ \sigma)^2$. We must reach at some moment the start of $\pi_i$ in our traversal of the word $(\pi_i \circ \sigma)^2$. But our underlying assumption implies that from here on, no word can fully traverse the corresponding $\pi_k$ (including $\pi_i$). Again, no word is fully traversed, as claimed. ∎

Lemma 15 yields immediately:

**Corollary 16** *Let $\pi_1, \pi_2, ..., \pi_p$ be a terminal collection of full word over some alphabet, and let $\pi_{p+1}, \pi_{p+2}, ..., \pi_n$ be arbitrary full words over the same alphabet. Then the collection*

$$(\pi_1 \circ \pi_2 \circ ... \circ \pi_n)^2, (\pi_1)^k, (\pi_2)^k, ..., (\pi_{i-1})^k, (\pi_{i+1})^k, ..., (\pi_p)^k$$

*is terminal as well. This holds for every $1 \leq i \leq p$ and $k \geq \sum_{i=1}^n |\pi_i|$.*

This is a special case of Lemma 15 where $\sigma = \pi_{i+1} \circ \ldots \pi_n \circ \pi_1 \ldots \circ \pi_{i-1}$.

## 3.2 The MC($n, 2n-1$) upper bound

The proof we present shows somewhat more than Theorem 1 says. We do this, since the scheduler can "trade" a player $P$ for a chair $c$. Namely, he can keep $P$ constantly on chair $c$. This allows the scheduler to move any other player past $c$-chairs. In other words this effectively means the elimination of chair $c$ from all other words. This suggests the following definition: If $\pi$ is a word over alphabet $C$ and $B \subseteq C$, we denote by $\pi(B)$ the word obtained from $\pi$ by deleting from it the letters from $C \setminus B$.

Our construction is recursive. An inductive step should add one player (i.e., a word) and two chairs. We carry out this step in two installments: In the first we add a single chair and in the second one we add a chair and a player. Both steps are accompanied by conditions that counter the above-mentioned trading option.

**Proposition 17** *For every integer $n \geq 1$*

- *There exist full words $s_1, s_2, ..., s_n$ over the alphabet $\{1, 2, ..., 2n-1\}$ such that $s_1(A), s_2(A), ..., s_p(A)$ is a terminal collection for every $p \leq n$, and every subset $A \subseteq \{1, 2, ..., 2n-1\}$ of cardinality $|A| = 2p - 1$.*

- *There exist full words $w_1, w_2, ..., w_n$ over alphabet $\{1, 2..., 2n\}$, such that $w_1(B), w_2(B), ..., w_p(B)$ is a terminal collection for every $p \leq n$, and every subset $B \subseteq \{1, 2, ..., 2n\}$ of cardinality $|B| = 2p - 1$.*

The words $s_1, s_2, ..., s_n$ in Proposition 17 constitute a terminal collection and are hence an oblivious $MC(n, 2n-1)$ algorithm that proves the upper bound part of Theorem 1. In the rest of this section we prove Proposition 17.

**Proof.**

As mentioned, the proof is by induction on $n$. For $n = 1$ clearly $s_1 = 11$ and $w_1 = 1122$ satisfy the conditions.

In the induction step we use the existence of $s_1, s_2, ..., s_n$ to construct $w_1, w_2, ..., w_n$. Likewise the construction of $s_1, s_2, ..., s_{n+1}$ builds on the existence of $w_1, w_2, ..., w_n$.

**The transition from** $w_1, w_2, ..., w_n$ **to** $s_1, s_2, ..., s_{n+1}$**:**

To simplify notations we assume that the words $w_1, w_2, ..., w_n$ in the alphabet $\{2, 3, ..., 2n+1\}$ (rather than $\{1, 2, ..., 2n\}$) satisfy the proposition. Let $k := \sum |w_i|$ and define:

$$\begin{aligned} s_1 &:= 1 \otimes ((w_1 \circ w_2 \circ ... \circ w_n)^{2(2n+1)}) \\ \forall i = 2, \ldots n+1 \quad s_i &:= (w_{i-1})^{k(2n+1)} \circ 1 \end{aligned}$$

Fix a subset $A \subseteq \{1, 2, ..., 2n+1\}$ of cardinality $|A| = 2p - 1$ with $p \leq n + 1$, and let us show that $s_1(A), s_2(A), ..., s_p(A)$ is a terminal collection. There are two cases to consider:

We first assume $1 \notin A$. This clearly implies that $p \leq n$ (or else $A = \{1, 2, ..., 2n+1\}$ and in particular $1 \in A$). In this case the collection is:

$$\begin{aligned} s_1(A) &:= ((w_1(A) \circ w_2(A) \circ ... \circ w_n(A))^{2(2n+1)}) \\ \forall i = 2, \ldots p \quad s_i(A) &:= (w_{i-1}(A))^{k(2n+1)} \end{aligned}$$

By the induction hypothesis, the collection $w_1(A), w_2(A), ..., w_{p-1}(A), w_p(A)$ is terminal. We apply Corollary 16 and conclude that

$$(w_1(A) \circ w_2(A) \circ ... \circ w_n(A))^2, (w_1(A))^k, (w_2(A))^k, ..., (w_{p-1}(A))^k$$

is terminal as well. But the $s_i$ are obtained by taking $(2n+1)$-th powers of these words, so that $s_1(A), s_2(A), ..., s_p(A)$ is terminal as needed.

We now consider what happens when $1 \in A$.

We define $F_1 := (w_1(A) \circ w_2(A) \circ ... \circ w_n(A))^2$ and for for $j > 1$, let $F_j := (w_{j-1}(A))^k$. We refer to $F_i$ as the $i$-th block. In our construction each word has $2n+1$ blocks, ignoring chair 1.

At any moment throughout a schedule we denote by $\mathcal{O}_1$ the set of players in $\{P_2, P_3, ..., P_p\}$ that currently occupy chair 1. We show that during a period in which the set $\mathcal{O}_1$ remains unchanged, no player can traverse a whole block. The proof splits according to whether $\mathcal{O}_1$ is empty or not.

Assume first that $\mathcal{O}_1 \neq \emptyset$, and pick some $i > 1$ for which $P_i$ occupies chair 1 during the current period. As long as $\mathcal{O}_1$ remains unchanged, $P_i$ stays on chair 1, so the words that the other players repeatedly traverse are as follows: For $P_1$ it is

$$w_1(A\backslash\{1\}) \circ w_2(A\backslash\{1\}) \circ ... \circ w_n(A\backslash\{1\})$$

and for $P_j$ with $p \geq j \neq i \geq 2$ it is

$$w_{j-1}(A\backslash\{1\})$$

We now show that no player can traverse a whole block (as defined above). Observe that the collection $\{w_\nu(A\backslash\{1\}) | \nu = 1, \ldots, p-1\}$ (including, in particular the word $w_{i-1}(A\backslash\{1\})$) is terminal. This follows from the induction hypothesis, because $|A\backslash\{1\}| = 2p - 2$, and because the property of being terminal is maintained under the insertion of new chairs into words. Applying Corollary 16 to this terminal collection implies that this collection of blocks is terminal as well.

We turn to consider the case $\mathcal{O}_1 = \emptyset$. In this case player 1 cannot advance from a none-1 chair to the next none-1 chair, since the two are separated by the presently unoccupied chair 1. We henceforth assume that player $P_1$ stays put on chair $c \neq 1$, but our considerations remain valid even if at some moment player $P_1$ moves to chair 1. (If this happens, he will necessarily stay there, since $\mathcal{O}_1 = \emptyset$). We are in a situation where players $P_2, P_3, ..., P_p$ traverse the words $w_1(A\backslash\{1,c\}), w_2(A\backslash\{1,c\}), ..., w_{p-1}(A\backslash\{1,c\})$ (chair $c$ which is occupied by player $P_1$ can be safely eliminated from these words). But $|A\backslash\{1,c\}| = 2p - 3$, so by the induction hypothesis no player can traverse a whole $w_i(A\backslash\{1,c\})$, so no player can traverse a whole block.

We just saw that during a period in which the set $\mathcal{O}_1$ remains unchanged, no player can traverse a whole block.

Finally, assume towards contradiction that $P_j$ fully traverses $s_j$ for some index $j$, and consider the first occurrence of such an event. It follows that $P_j$ has traversed $2n+1$ blocks, so that the set $\mathcal{O}_1$ must have changed at least $2n+1$ times during the process. However, for $\mathcal{O}_1$ to change, some $P_i$ must either move to, or away from a 1-chair in $s_i$. But 1 occurs exactly once in $s_i$, so every $P_i$ can account for at most two changes in $\mathcal{O}_1$, a contradiction.

**The transition from $s_1, s_2, ..., s_n$ to $w_1, w_2, ..., w_n$:**

We assume that the words $s_1, s_2, ..., s_n$ in the alphabet $\{2, 3, ..., 2n\}$ satisfy the proposition. Let $k := \sum |s_i|$ and define:

$$w_1 : = 1 \otimes ((s_1 \circ s_2 \circ ... \circ s_n)^{2(2n+1)})$$
$$\forall i = 2, \ldots, n \quad w_i : = (s_{i-1})^{k(2n+1)} \circ 1$$

Fix a subset $B \subseteq \{1, 2, ..., 2n\}$ with $|B| = 2p - 1$. Then

$$w_1(B) = 1 \otimes ((s_1(B) \circ s_2(B) \circ ... \circ s_n(B))^{2(2n+1)})$$
$$\forall i = 2, \ldots, p \quad w_i(B) = (s_{i-1}(B))^{k(2n+1)} \circ 1$$

are exactly the same as in the previous transition just by replacing $s$ with $w$ and $A$ with $B$ (in this case the induction hypothesis is on $s_i$ and we prove for $w_i$). So exactly the same considerations prove that $w_1(B), w_2(B), ..., w_m(B)$ is a terminal collection. ∎

## 4 The oblivious Renaming$(n, 2n - 1)$ algorithm

The ideas developed to solve the musical chairs problem and prove Theorem 1 turn out to yield as well an answer to the oblivious Renaming problem and a proof of Theorem 2. The rules are the same as in the $MC$ problem, except that the scheduler cannot select the initial positions, and every word is started at its first letter. In order to prove Theorem 2 we should construct a collection of full words $\Pi_N = \{s_1, s_2, ..., s_N\}$ over the alphabet $[2N - 1]$ such that for every $n \leq N$ and for every set of $n$ words from $\Pi_N$ the following holds: Every schedule that starts from the first letter in each of these words reaches a safe configuration and all players only visits chairs from the set $\{1, \ldots, 2n - 1\}$.

We note that our construction yields very long words - triply exponential in $N$. It is an interesting challenge to accomplish this with substantially shorter words.

**Proof.**[Theorem 2] By Proposition 14 and Theorem 1, we can construct for each $1 \leq i, n \leq N$ a word $\pi_{i,n}$ that is $[2n-1]$-full such that every set of $n$ words in the set $\{\pi_{i,n} | i = 1, \ldots, N\}$ constitute an oblivious $MC(n, 2n - 1)$ protocol.

We show that with a proper choice of the exponents $l_1, \ldots, l_N$, the Theorem holds with the words $s_i = \pi_{i,1}^{l_1} \circ \pi_{i,2}^{l_2} \circ \ldots \circ \pi_{i,N}^{l_N}$.

The theorem follows if we can show that for every $1 \leq n \leq N$ and every subset $J \subseteq [N]$ of cardinality $|J| = n$ the following holds: In every possible schedule that starts each word in $\{s_j | j \in J\}$ from its first letter, no player reaches a position beyond the subword $\pi_{j,n}^{l_n}$. Consider any point in such a schedule. Say that player $P_j$ (for some $j \in J$) is *leading* if it currently resides in the stretch $\pi_{j,n}^{l_n}$ of $s_j$. Otherwise, we say that $j$ is *trailing*. We observe that during a period of time in which no trailing player changes position, no leading player can traverse a complete copy of $\pi_{j,n}$. To see this, consider an arbitrary $MC$ schedule with the words $\{\pi_{j,n} | j \in J\}$. We start this schedule as follows: Every leading player maintains his position from the original renaming schedule and every trailing player stays put on the same chair that he is currently occupying. (Such a chair can be found in the word $\pi_{j,n}$ since it is $[2n-1]$-full). The claim follows since the words $\{\pi_{j,n} | j \in J\}$ constitute an oblivious $MC(n, 2n - 1)$ protocol.

It follows that no leading player $P_j$ can traverse more than $\sum_{\nu<n, i \in J\setminus\{j\}} |\pi_{i,\nu}| l_\nu$ copies of $\pi_{j,n}$ in $s_j$. Our claim follows if we choose $l_j$ that is larger than this integer. ∎

# 5 Oblivious MC algorithms via the probabilistic method

We start with an observation that puts Theorems 3 and 4 (as well as Theorem 1) in an interesting perspective. The expected number of pairwise collisions in a random configuration is exactly $\binom{n}{2}/m$. In particular, when $m \gg n^2$, most configurations are *safe* (namely, have no collisions). Therefore, it in not surprising that in this range of parameters $n$ random words would yield an oblivious $MC(n, m)$ algorithm. However, when $m = O(n)$, only an exponentially small fraction of configurations are safe, and the existence of oblivious $MC(n, m)$ algorithms is far from obvious.

## 5.1 Full words with $O(n)$ chairs, allowing repetitions

Theorem 3 can be thought of as a (nonconstructive) derandomization of the randomized MC algorithm in which players choose their next chair at random (and future random decisions of players are not accessible to the scheduler). Standard techniques for derandomizing random processes involve taking a union bound over all possible bad events, which in our case corresponds to a union bound over all possible schedules. The asynchronous scheduler has too many options (and so does the immediate scheduler), making a union bound too wasteful. For this reason, we shall consider in this section the canonical scheduler, which is as powerful as the asynchronous scheduler (see Section 2.5). In every unsafe configuration, the choice of canonical pair is deterministic and the canonical scheduler has only three possible moves to choose from, which makes it viable to use a union bound. We now prove Theorem 3.

**Proof.** Each of the $N$ words is chosen independently at random as a sequence of $L$ chairs, where each chair in the sequence is chosen independently at random. We show that with high probability (probability tending to 1 as the value of the constant $c$ grows), this choice satisfies Theorem 3.

It is easy to verify that in this random construction, with high probability, all words are full. To see this note that the probability that chair $j$ is missing from word $i$ is $((m-1)/m)^L$. Consequently, the probability that a word chosen this way is not full is $\leq m((m-1)/m)^L$. Therefore, the expected number of non-full words is $\leq m \cdot N \cdot ((m-1)/m)^L$. But with our choice of parameters $m = 7n$ and $L = cn \log N$, we see that $m \cdot N \cdot ((m-1)/m)^L = o(1)$, provided that $c$ is large enough.

In our approach to the proof we keep track of all possible schedules. To this end we use "a logbook" that is the complete ternary tree $\mathcal{T}$ of depth $L$ rooted at $r$. Associated with every node $v$ of $\mathcal{T}$ is a random variable $X_v$. The values taken by $X_v$ are system configurations. For a given choice of words and an initial system configuration we define the value of $X_r$ to be the chosen initial configuration. Every node $v$ has three children corresponding to the three possible next configurations that are available to the canonical scheduler at configuration $X_v$.

Another important ingredient of the proof is a *potential* function (defined below) that maps system configurations to the nonnegative reals. It is also convenient to define an (artificial) "empty" configuration of 0 potential. Every safe configuration has potential 1, and every non-empty unsafe configuration has potential $> 10$. If the node $u$ is a descendant of $v$ and the system configuration $X_v$ is safe, then we define $X_u$ to be the empty configuration.

We thus also associate with every node of $\mathcal{T}$ a nonnegative random variable $P = P_v$ that is is the potential of the (random) configuration $X_v$. The main step of the proof is to show that if $v_1, v_2, v_3$ are the three children of $v$, then $\sum_{i=1}^{3} \mathbb{E}(P_{v_i}) \leq r\mathbb{E}(P_v)$ for some constant $r \leq 0.99$. (Note that this inequality holds as well if $X_v$ is either safe or empty). This exponential drop implies that

$$\mathbb{E}(\sum_{v \text{ is a leaf of } \mathcal{T}} (P_v)) = \sum_{v \text{ is a leaf of } \mathcal{T}} \mathbb{E}(P_v) = o(1)$$

provided that $L$ is large enough. This implies that with probability $1 - o(1)$ (over the choice of random words) all leaves of $\mathcal{T}$ correspond to an empty configuration. In other words every schedule terminates in fewer than $L$ steps.

We turn to the details of the proof. A configuration with $i$ occupied chairs is defined to have potential $x^{n-i}$, where $x > 1$ is a constant to be chosen later. In a nonempty configuration the potential can vary between 1 and $x^{n-1}$, and it equals 1 iff the configuration is safe.

Consider a configuration of potential $x^{n-i}$ (with $i < n$), where the canonical pair is $(\alpha, \beta)$. It has three children representing the move of either $\alpha$ or $\beta$ or both. Let us denote $\rho = i/m$ and $\rho' = (i-1)/m$. When a single player moves, the number of occupied chairs can stay unchanged, which happens with probability $\rho$. With probability $1 - \rho$ one more chair will be occupied and the potential gets divided by $x$. Consider next what happens when both players move. Here the possible outcomes (in terms of number of occupied chairs) depend on whether there is an additional player $\gamma$ currently co-occupying the same chair as $\alpha$ and $\beta$. It suffices to perform the analysis in the less favorable case in which there is no such player $\gamma$, as this provides an upper bound on the potential also for the case that there is such a player. With probability $(\rho')^2$ both $\alpha$ and $\beta$ move to occupied chairs and the potential gets multiplied by $x$. With probability $\rho'(1-\rho') + (1-\rho')\rho = (\rho+\rho')(1-\rho')$ the number of occupied chairs (and hence the potential) does not change. With probability $(1-\rho')(1-\rho)$ the number of occupied chairs grows by one and the potential gets divided by $x$.

It follows that if $v$ is a node of $\mathcal{T}$ with children $v_1, v_2, v_3$ and if the configuration $X_v$ is unsafe and nonempty then $\sum_{i=1}^{3} \mathbb{E}(P_{v_i}) \leq \mathbb{E}(P_v)(2\rho + 2(1-\rho)/x + (\rho')^2 x + (\rho+\rho')(1-\rho') + (1-\rho)(1-\rho')/x)$. Recall that $x > 1$ and $\rho' < \rho < 1$. This implies that the last expression increases if $\rho'$ is replaced by $\rho$, and thereafter it is maximized when $\rho$ attains its largest possible value $q = (n-1)/m$. We conclude that

$$\sum_{1}^{3} \mathbb{E}(P_{v_i}) \leq \mathbb{E}(P)(2q + 2(1-q)/x + q^2 x + 2q(1-q) + (1-q)^2/x).$$

We can choose $q = 1/7$ and $x = 23/2$ to obtain $\sum_{i=1}^{3} \mathbb{E}(P_{v_i}) \leq r\mathbb{E}(P_v)$ for $r < 0.99$. This guarantees an exponential decrease in the expected sum of potentials and hence termination, as we now explain.

It follows that for every initial configuration the expected sum of potentials of all leaves at depth $L$ does not exceed $x^{n-1}$ (the largest possible potential) times $r^L$. On the other hand, if there is at least one leaf $v$ for which the configuration $X_v$ is neither safe nor empty, then the sum of potentials at depth $L$ is at least $x > 1$. Our aim is to show that with high probability (over the choice of $N$ words), all runs have length $< L$: (i) For every choice of $n$ out of the $N$ words, (ii) Each selection of an initial configuration, and (iii) Every canonical scheduler's strategy. The $n$ words can be chosen in $\binom{N}{n}$ ways. For every $n$ words, there are $L^n$ possible initial configurations. The probability of length-$L$ run from a given configuration is at most $x^{n-1}r^L$, where $x = 23/2$ and $r < 0.99$. Therefore our claim is proved if $\binom{N}{n} \cdot x^{n-1} r^L \leq o(1)$. This inequality clearly holds if we let $L = cn \log N$ with $c$ a sufficiently large constant. This completes the proof of Theorem 3. ∎

A careful analysis of the proof of Theorem 3 shows that it actually works as long as $\frac{m}{n} > 4 + 2\sqrt{2} = 6.828...$ It would be interesting to determine the value of $\liminf_{n\to\infty} \frac{m}{n}$ for which $n$ long enough random words over an $m$-letter alphabet constitute, with high probability, an oblivious $MC(n,m)$ protocol.

## 5.2 Permutations over $O(n)$ chairs

The argument we used to prove Theorem 3 is inappropriate for the proof of Theorem 4. Theorem 4 deals with random permutations, whereas in the proof of Theorem 3 we use words of length $\Omega(n \log n)$. (Longer words are crucial there for two main reasons: To guarantee that words are full and to avoid wrap-around. The latter property is needed to guarantee independence.) Indeed in proving Theorem 4 our arguments are substantially different. In particular, we work with a pairwise immediate scheduler, and unlike the proof of Theorem 3, there does not appear to be any significant benefit (e.g., no significant reduction in the ratio $\frac{m}{n}$) if a canonical scheduler is used instead.

We first prove the special case $N = n$ of Theorem 4.

**Theorem 18** *If $m \geq cn$ where $c > 0$ is a sufficiently large constant, then there is a family of $n$ permutations on $[m]$ which constitute an oblivious $MC(n, m)$ protocol.*

We actually show that with high probability, a set of random permutations $\pi_1, \ldots, \pi_n$ has the property that in every possible schedule the players visit at most $L = O(m \log m)$ chairs. Our analysis uses the approach of deferring random decisions until they are actually needed. For each of the $m^n$ possible initial configuration, we consider all possible sequences of $L$ locations. For each such sequence we fill in the chairs in the locations in the sequence at random, and prove that the probability that this sequence represents a possible schedule is extremely small – so small that even if we take a union bound over all initial configurations and over all sequences of length $L$, we are left with a probability much smaller than 1.

The main difficulty in the proof is that since $L \gg m$ some players may completely traverse their permutation (even more than once) and therefore the chairs in these locations are no longer random. To address this, we partition the sequence of moves into $L/t$ blocks, where in each block players visit a total of $t$ locations. We can and will assume that $t$ divides $L$. We take $t = \delta m$ for some sufficiently small constant $\delta$, and $n = \epsilon m$, where $\epsilon$ is a constant much smaller than $\delta$. This choice of parameters implies that within a block, chairs are essentially random and independent. To deal with dependencies among different blocks, we classify players (and their corresponding permutations) as *light* or *heavy*. A player is *light* if during the whole schedule (of length $L$) it visits at most $t/\log m = o(t)$ locations. A player that visits more than $t/\log m$ locations during the whole sequence is *heavy*. Observe that for light players, the probability of encountering a particular chair in some given location is at most $\frac{1}{m-o(t)} \leq \frac{1+o(1)}{m}$. Hence, the chairs encountered by light players are essentially random and independent (up to negligible error terms). Thus it is the heavy players that introduce dependencies among blocks. Every heavy player visits at least $t/\log m$ locations, so that $n_h$, the number of heavy players does not exceed $n_h \leq (L \log m)/t = O(\log^2 m)$. The fact that the number of heavy players is small is used in our proof to limit the dependencies among blocks.

The following lemma is used to show that in every block of length $t$ the number of locations that are visited by heavy players is not too large. Consequently, sufficiently many locations are visited by light players. In the lemma we use the following notation. A segment of $k$ locations in a permutation is said to have *volume* $k - 1$. Given a collection of locations, a chair is *unique* if it appears exactly once in these locations.

**Lemma 19** *Let $n_h \leq m/\log^2 m$ and let $\delta > 0$ be a sufficiently small constant. Consider $n$ random permutations over $[m]$. Select any $n_h$ of the permutations and a starting location in each of them. Choose next intervals in the selected permutations with total volume $t'$ for some $t/10 \leq t' \leq t$. With probability $1 - o(1)$ for every such set of choices at least $4t'/5$ of the chairs in the chosen intervals are unique.*

**Proof.** We first note that we will be using the lemma with $n_h = O(\log^2 n)$. Also, if a list of letters contains $u$ unique letters (i.e., they appear exactly once) and $r$ repeated letter (i.e., appearing at least twice), then it has $d = u+r$ distinct letters and length $\lambda \geq u+2r$. In particular $d \leq (\lambda+u)/2$.

There are $\binom{n}{n_h}$ ways of choosing $n_h$ of the permutations. Then, there are $m^{n_h}$ choices for the initial configuration. We denote by $s_i$ the volume of the $i$-th interval, so that $\sum_{i=1}^{n_h} s_i = t'$. Therefore there are $\binom{t'+n_h-1}{n_h-1} \leq m^{n_h}$ ways of choosing the intervals with total volume $t'$. Since the volume of every interval is at most $t'$ we have that the probability that a particular chair resides at a particular location in this interval is at most $1/(m-t')$. This is because the permutation is random and at most $t'$ chairs appeared so far in this interval. Therefore the probability that a sequence of $t'$ labels involves less than $0.9t'$ distinct chairs is at most

$$\binom{m}{0.9t'}\left(\frac{0.9t'}{m-t'}\right)^{t'} \leq \left(\frac{em}{0.9t'}\right)^{0.9t'}\left(\frac{0.9t'}{m-t'}\right)^{t'} \leq e^{t'}\left(\frac{m}{m-t'}\right)^{0.9t'}\left(\frac{t'}{m-t'}\right)^{0.1t'}$$
$$\leq 4^{t'}(2\delta)^{0.1t'} \ll e^{-t'}.$$

Explanation: The set of chairs that appear in these intervals can be chosen in $\binom{m}{0.9t'}$ ways. The probability that a particular location in this union of intervals is assigned to a chair from the chosen set does not exceed $\frac{0.9t'}{m-t'}$. In addition $m/(m-t') \leq (1+\delta)$, $t'/(m-t') \leq 2\delta$ and $\delta$ is a very small constant.

Now we take a union bound over all choices of $n_h$ permutations, all starting locations and all collection of intervals with total volume $t'$. It follows that the probability that there is a choice of intervals of volume $t'$ that span $\leq n_h$ permutations and contain fewer than $9t'/10$ distinct chairs is at most

$$m^{3n_h}e^{-t'} = o(1).$$

In the above notation $\lambda = t'$ and $d \geq 0.9t'$ which yields $u \geq 0.8t'$ as claimed. ∎

Since the conclusion of this lemma holds with probability $1 - o(1)$ we can assume that our set of permutations satisfies it. In particular, in every collection of intervals in these permutations with total volume $\frac{t}{10} \leq t' \leq t$ that reside in $O(\log^2 m)$ permutations there are at least $4t'/5$ unique chairs.

As already mentioned, we break the sequence of $L$ locations visited by players into blocks of $t$ locations each. We analyze the possible runs by considering first the *breakpoints profile*, namely where each block starts and ends on each of the $n$ words. There are $m^n$ possible choices for the starting locations. If, in a particular block player $i$ visits $s_i$ chairs, then $\sum_{i=1}^n s_i = t$. Consequently the parameters $s_1, \ldots, s_n$ can be chosen in $\binom{t+n-1}{n} \leq 2^{t+n}$ ways. There are $L/t$ blocks, so that the total number of possible breakpoints profiles is at most $m^n(2^{t+n})^{L/t} \leq m^n 2^{2L}$ (here we used the fact that $t > n$). Clearly, by observing the breakpoints profile we can tell which players are light and which are heavy. We recall that there are at most $O(\log^2 m)$ heavy players, and that the premise of Lemma 19 can be assumed to hold.

Let us fix an arbitrary particular breakpoints profile $\beta$. We wish to estimate the probability (over the random choice of chairs) that some legal sequence of moves by the pairwise immediate scheduler yields this breakpoints profile $\beta$. Let $B$ be an arbitrary block in $\beta$. Let $p(B)$ denote the probability over choice of random chairs and *conditioned over contents of all previous blocks in $\beta$* that there is a legal sequence of moves by the pairwise immediate scheduler that produces this block $B$.

**Lemma 20** *For $p(B)$ as defined above we have that $p(B) \leq 8^{-t}$.*

**Proof.** The total number of chairs encountered in block $B$ is $n \ll t$ (for the initial locations) plus $t$ (for the moves). Recall that the set of heavy players is determined by the block-sequence $\beta$. Hence within block $B$ it is clear which are the heavy players and which are the light players. Let $t_h$ (resp. $t_\ell = t - t_h$) be the number of chairs visited by heavy (resp. light) players in this block. The proof now breaks into two cases, depending on the value of $t_h$.

**Case 1:** $t_h \leq 0.1t$. Light players altogether visit $n + t_\ell$ chairs ($n$ initial locations plus $t_\ell$ moves). If $u$ of these chair are unique, then they visit at most $(n + t_\ell + u)/2$ distinct chairs. But a chair in this collection that is unique is either: (i) One of the $n$ chairs where a player terminates his walk, or, (ii) A chair that a light player traverses due to a collision with a heavy player, and there are at most $t_h$ of those. Consequently, the number of distinct chairs visited by light players does not exceed $(n + t_\ell + n + t_h)/2 = t/2 + n$.

Fix the set $S$ of $t/2 + n$ distinct chairs that we are allowed to use. There are $\binom{m}{n+t/2}$ choices for $S$. Now assign chairs to the locations one by one, in an arbitrary order. Each location has probability of at most $(1 + o(1))\frac{n+t/2}{m}$ of receiving a chair in $S$. Since we are dealing here with light players, we have exposed only $o(m)$ chairs for each of them (in $B$ and in previous blocks of $\beta$), and as mentioned above, this can increase the probability by no more that a $1 + o(1)$ factor.

Hence the probability that the segments traversed by the light players contain only $n + t/2$ chairs is at most

$$\binom{m}{n+t/2}\left((1+o(1))\frac{n+t/2}{m}\right)^{t_\ell} \leq \left(\frac{em}{n+t/2}\right)^{n+t/2} 2^{t_\ell}\left(\frac{n+t/2}{m}\right)^{t_\ell}$$

$$\leq (2e)^t \left(\frac{n+t/2}{m}\right)^{(t_\ell - t_h)/2 - n} \leq (2e)^t (t/m)^{t/4} < 8^{-t}.$$

Here we used that $t_h + t_\ell = t$, $t_h \leq 0.1t$, $t_l \geq 0.9t$ and $n \ll t \ll m$.

**Case 2:** $t_h \geq 0.1t$. Let us reveal first the chairs visited by the heavy players. By Lemma 19, we find there at least $4t_h/5$ unique chairs. In order that the heavy players traverse these chairs, they must be visited by light players as well. Hence the $t_\ell$ locations visited by light players must include all these $0.8t_h$ pre-specified chairs. We bound the probability of this as follows. First choose for each of the $0.8t_h$ pre-specified chairs a particular location where it should appear in the intervals of light players. The number of such choices is $\leq t_\ell^{0.8t_h}$. As mentioned above the probability that a particular chair is assigned to some specific location is $(1+o(1))/m$. Therefore the probability that $0.8t_h$ pre-specified chairs appear in the light intervals is at most $t_\ell^{0.8t_h}((1+o(1))/m)^{0.8t_h}$. Thus the probability that a schedule satisfying the condition of the lemma exists is at most

$$t_\ell^{0.8t_h}((1+o(1))/m)^{0.8t_h} \leq (2t/m)^{0.8t_h} \leq (2t/m)^{t/15} < 8^{-t},$$

where we used that $n \ll t \ll m$. ∎

Lemma 20 implies an upper bound of $p(B)^{L/t} = 8^{-L}$ on the probability there is a legal sequence of moves by the pairwise immediate scheduler that gives rise to breakpoints profile $\beta$. Taking a union bound over all block sequences (whose number is at most $m^n 2^{2L} \leq 6^L$, by our choice of $L = Cm \log m$ for a sufficiently large constant $C$), Theorem 18 is proved.

Observe that the proof of Theorem 18 easily extends to the case that there are $N = m^{O(1)}$ random permutations out of which one chooses $n$. We simply need to multiply the number of possibilities by $N^n$, a term that can be absorbed by increasing $m$, similar to the way the term $m^n$ is absorbed. In Lemma 19 we need to replace $\binom{n}{n_h}$ by $\binom{N}{n_h}$, and the proof goes through without any change (because $n_h$ is so small). This proves Theorem 4.

## 5.3 Explicit construction with permutations and $m = O(n^2)$

In this section we present for every integer $d \geq 1$ an explicit collection of $n^d$ permutations on $m = O(d^2 n^2)$ such that every $n$ of these permutations constitute an oblivious $MC(n, m)$ algorithm. This proves Theorem 5.

We let $LCS(\pi, \sigma)$ stand for the length of the longest common subsequence of the two permutations $\pi$ and $\sigma$, considered cyclically. (That is, we may rotate $\pi$ and $\sigma$ arbitrarily to maximize the length of the resulting longest common subsequence). The following easy claim is useful.

**Proposition 21** *Let $\pi_1, \ldots, \pi_n$ be permutations of $\{1, \ldots, m\}$ such that $LCS(\pi_i, \pi_j) \leq r$ for all $i \neq j$. If $m > (n-1)r$, then in every schedule none of the $\pi_i$ is fully traversed.*

**Proof.** By contradiction. Consider a schedule in which one of the permutations is fully traversed, say that $\pi_1$ is the first permutation to be fully traversed. Each move along $\pi_1$ reflects a collision with some other permutation. Hence there is a permutation $\pi_i, i > 1$ that has at least $m/(n-1)$ agreements with $\pi_1$. Consequently, $r \geq LCS(\pi_1, \pi_i) \geq \frac{m}{(n-1)}$, a contradiction. ∎

This yields an inexplicit oblivious $MC(n, m)$ algorithm with $m = O(n^2)$, since (even exponentially) large families of permutations in $[m]$ exist where every two permutations have an LCS of only $O(\sqrt{m})$. We omit the easy details. On the other hand, we should notice that by [4] this approach is inherently limited and can, at best yield bounds of the form $m \leq O(n^{3/2})$.

We now present an explicit construction that uses some algebra.

**Lemma 22** *Let $p$ be a prime power, let $d$ be a positive integer and let $m = p^2$. Then there is an explicit family of $(1 - o(1))m^d$ permutations of an $m$-element set, where the LCS of every two permutations is at most $4d\sqrt{m}$.*

**Proof.** Let $\mathbb{F}$ be the finite field of order $p$. Let $\mathcal{M} := \mathbb{F} \times \mathbb{F}$, and $m = p^2 = |\mathcal{M}|$. Let $f$ be a polynomial of degree $2d$ over $\mathbb{F}$ with vanishing constant term, and let $j \in \mathbb{F}$. We call the set $B_{f,j} = \{(x, f(x) + j) | x \in \mathbb{F}\}$ *a block*. We associate with $f$ the following permutation $\pi_f$ of $\mathcal{M}$: It starts with an arbitrary ordering of the elements in $B_{f,0}$ followed by $B_{f,1}$ arbitrarily ordered, then of $B_{f,2}$ etc. A polynomial of degree $r$ over a field has at most $r$ roots. It follows that for every two polynomials $f \neq g$ as above and any $i, j \in \mathbb{F}$, the blocks $B_{f,i}$ and $B_{g,j}$ have at most $2d$ elements in common. There are $(p-1) \cdot p^{2d-1} = (1 - o(1))m^d$ such polynomials. There are $p$ blocks in $\pi_f$ and in $\pi_g$, so that $LCS(\pi_f, \pi_g) \leq 4dp$, as claimed. ∎

## 6 Discussion and Open Problems

In this paper we introduced the notion of oblivious distributed algorithms. Our main results concern the design of oblivious MC algorithms. We showed that $m \geq 2n - 1$ chairs are necessary and sufficient for the existence of an oblivious $MC$ algorithm with $n$ processors. However, our construction involves very long words. It is interesting to find explicit constructions with $m = 2n - 1$ chairs and substantially shorter words.

In other ranges of the problem we can show, using the probabilistic method, that oblivious $MC(n, m)$ algorithms exist with $m = O(n)$ and relatively short full words. We still do not have explicit constructions of such protocols. We would also like to determine $\liminf \frac{m}{n}$ such that $n$ random words over an $m$ letter alphabet tend to constitute an oblivious $MC(n, m)$ algorithm.

Computer simulations strongly suggest that for random permutations, a value of $m = 2n - 1$ does not suffice. On the other hand, we have constructed (details omitted from this manuscript) oblivious $MC(n, 2n - 1)$ algorithms using permutations for $n = 3$ and $n = 4$ (for the latter the proof of correctness is computer-assisted). For $n \geq 5$ we have neither been able to find such systems (not even in a fairly extensive computer search) nor to rule out their existence.

A self contained proof of the $m \geq 2n - 1$ lower bound will appear in a subsequent paper. The following question remains open: What is the smallest $m$ for which there are collections of $N = m + 1$ (not necessarily full) words such that every $\min[n, N]$ of them form an oblivious $MC$ algorithm *when starting at the initial chair of each word*. Our proof that $m \geq 2n - 1$ assumes that the scheduler is allowed to pick an arbitrary initial state on each word.

We do not know how hard it is to recognize whether a given collection of words constitute an oblivious $MC$ algorithm. This can be viewed as the problem whether some digraph contains a directed cycle or not. The point is that the digraph is presented in a very compact form. It is not hard to place this problem in PSPACE, but is it in a lower complexity class, such as co-NP or P?

There are interesting foundational questions related to different models in distributed computing. We have defined here the Output Negotiation ($ON$) model, and showed that it is properly included in the read/write model. It follows by definition that the oblivious model is included in the $ON$ model. It would be interesting to know whether this last inclusion is proper.